\documentclass[journal]{IEEEtran}

\usepackage{color,soul}
\usepackage{xcolor}
\colorlet{correction1}{black}

\usepackage{cite}

\ifCLASSINFOpdf
\else
\fi
\usepackage{amsmath}
\usepackage{amsfonts}
\usepackage{algorithm}
\usepackage[noend]{algpseudocode}
\colorlet{correction2}{black}
\makeatletter
\def\BState{\State\hskip-\ALG@thistlm}
\makeatother
\usepackage{graphicx}

\usepackage[caption=false,font=normalsize,labelfont=sf,textfont=sf]{subfig}

\usepackage{url}

\usepackage[nolist,nohyperlinks]{acronym}

\begin{document}
%
\title{Minimizing the Signaling Overhead and Latency based on Users' Mobility Patterns}

\author{Merim Dzaferagic, Nicola Marchetti, Irene Macaluso, \\
\{dzaferam, nicola.marchetti, macalusi\}@tcd.ie, CONNECT, Trinity College Dublin, Ireland \vspace{-2.5em}}


%


\maketitle

\begin{abstract}
We demonstrate a distributed and a centralized 4G/5G compliant approach to minimize signaling and latency related to user mobility in cellular networks. \textcolor{correction2}{This is crucial due to the densification of networks and the additional signaling introduced by the new 5G service based architecture.} By exploiting standardized protocols, our solutions dynamically reorganize the association between nodes in Radio Access Network (RAN) and the core. We validated the proposed approaches using real user mobility datasets. Our results show that both our distributed and centralized solutions significantly reduce the signaling between core and RAN compared to the traditional approach based on geographical proximity. \textcolor{correction2}{As a result, both approaches  significantly reduce the average handover procedure processing time.} Moreover, by relying on locally available information, the distributed approach can quickly adapt to changes in the user movement patterns as they happen.

\end{abstract}


%
\IEEEpeerreviewmaketitle


\section{Introduction}\label{sec:introduction}


Due to the increasing densification of  cellular networks and mobile devices (e.g. Internet of Things devices for logistics and supply chain management), the number of handovers will significantly grow. \textcolor{correction1}{Mobility management optimization based on \ac{UE} mobility patterns recognition has been identified as one of the key issues by the \ac{3GPP} \cite{3gpp_2018_16}.}
Not all handovers come with the same cost in terms of signaling to the core network and therefore in terms of latency. \textcolor{correction2}{One of the key factors to determine the signalling overhead associated to handover procedures is the association between nodes in the \ac{RAN} and nodes
 (e.g. \ac{MME} in the case of 4G),
functions
 (e.g. \ac{AMF} in the case of 5G),
and Location Regions 
(e.g. \ac{TA}, Registration Area, \ac{TA} List)
in the core network. The focus of this work is on this \ac{RAN}-to-core association for 4G, i.e. \ac{RAN}-to-\ac{MME} association, and 5G, i.e. \ac{RAN}-to-\ac{AMF} association.}  
We focus on both 4G and 5G for two reasons: the \ac{RAN}-to-core association is similar in both architectures, and  the transition from 4G to 5G will be evolutionary rather than revolutionary meaning that both technologies will coexist for a time \cite{patzold20185g}.
 

 \textcolor{correction2}{In the remainder of this paper we refer to handovers requiring an \ac{MME}/\ac{AMF} reallocation as inter-region handovers. This is the most inefficient type of handover in terms of delay and number of exchanged signaling messages. In fact, inter-region handovers require on average 50\% more signaling messages compared to the intra-region ones, i.e. handovers that do not result in a change of the \ac{MME}/\ac{AMF} \cite{3gpp_2007_12}. That results in higher latency of the inter-region handover procedures compared to the intra-region handovers \cite{3gpp_2011}. In this paper we propose an approach to reorganize nodes in the network so that the number of handovers requiring an \ac{MME} reallocation in the case of 4G or an \ac{AMF} reallocation in the case of 5G is minimized.} Our  smart design of the handover regions based on \ac{UE} mobility  information can be implemented in a  distributed or centralized manner. The distributed node re-configuration relies on existing protocol messages and network management systems \cite{3gpp_2008_36_423, 3gpp_28_533_v_15_r_15}. The centralized mechanism also relies on the same information, but due to the lack of requirements for these messages to be forwarded to the \ac{OSS}, it has to be implemented as a passive probe in the core network. The computational complexity of the distributed solution is $O(1)$ as compared to centralized one that is an NP-hard problem.
Both approaches  balance the load between the \ac{MME}/\ac{AMF} instances and are independent of the specific handover mechanisms implemented in the network. 
Moreover, they successfully adjust to both, the up- and down-scaling of resources in the core network (e.g. new \ac{AMF} instance) and the changes in \ac{UE} mobility patterns in the \ac{RAN}.

\textcolor{correction2}{The reminder of the paper is organized as follows. Section \ref{sec:related_work} surveys the related work. Section \ref{sec:solution} details the handover optimization problem and the proposed solutions. Section \ref{sec:performance_evaluation} presents the evaluation of the proposed solutions. Finally, Section \ref{sec:conclusions} concludes the paper. }

\section{Related Works}\label{sec:related_work}

\textcolor{correction2}{In the literature, a number of approaches have been proposed for autonomic network optimization. The authors of \cite{sebastian2018anticipatory, alawe2018improving, alawe2018smart, alawe2018scalability} highlight the existing and anticipate the upcoming problems related to signaling traffic, which will overload the \ac{AMF} instances. In addition, authors of \cite{sodhro2019quality} call attention to issues related to quality of service and high mobility. All of them agree that the flexibility of \acp{VNF} and intelligent resource allocation will be the key ingredient to solve these problems. The work in \cite{sebastian2018anticipatory}, focuses on the exploitation of the prediction of user behavior to improve post-handover procedures. The authors of \cite{alawe2018improving, alawe2018smart} show preliminary results proving that traffic forecasting techniques based on machine learning outperform the threshold-based solutions for dynamic up- and down-scaling of network resources. Similarly, in \cite{alawe2018scalability}, the authors propose a control theory based algorithm for autonomic up- and down-scaling of \acp{AMF} in the network. Our work also focuses on the optimization of the signaling overhead and the control plane latency, but instead of forecasting the traffic load or performing the up- and down-scaling of resources, we exploit information on user mobility to perform intelligent re-configurations of the network nodes.}

\textcolor{correction2}{In \cite{kunz2010minimizing} a mechanism to minimize the number of \ac{SGW} relocations is proposed. While \ac{MME} relocation is also mentioned, the authors focus on the \ac{SGW}. They propose the introduction of a Service Area for idle users, which is a subset of the Service Area of each \ac{SGW}. Although the results show a decrease of the number of \ac{SGW} relocations for users in active mode, the mechanism to determine the idle-mode service areas is not provided and the implications in terms of existing standards are not discussed. In \cite{taleb2013gateway} a centralized heuristic mechanism to deploy network functions so as to minimize the number of \ac{SGW} relocations is proposed. The minimization of \ac{MME} relocations is only mentioned as a possible application. However, the proposed solution relies on a greedy algorithm with high computational complexity (i.e.~$O(N^3)$) that does not assure a significant improvement in terms of \ac{SGW} relocations. Moreover, the authors do not discuss the improvements compared to the existing industry standard techniques for network planning and they do not provide details about the integration of their solution with existing or future networks. In contrast, our solution  exploits standardized protocols and mechanisms, which will facilitate its adoption.}

\textcolor{correction2}{Another approach in literature for inter-region handover optimization relies on a new architecture, proposed in \cite{moradi2014softmow}. The authors propose a recursive hierarchical algorithm to minimize the number of inter-region handovers in \cite{moradi2014softmow2}. In contrast, the solution we propose is designed to work within existing (4G) and emerging (5G) mobile networks so as to maximize technology adoption. Specific solutions for handover latency reduction have been proposed in the case of femto cells \cite{qian2014optimized, xu2011apparatus}. These two patents detail an optimized intra-HeNB GW (Home eNodeB Gateway) handover mechanism that reduces signaling to and from an \ac{MME}. While our approach also aims at reducing the signaling to and from the \ac{MME}, our network-wide approach is not restricted to the specific femto cell case and we take into account users' mobility when optimizing the nodes configuration. In \cite{velamati2015system, chowdhury2013dynamic}, the inventors detail a mechanism to re-assign \acp{BS} to \acp{MME} so as to balance the load of the \acp{MME}. Our solution, while also considering the load, re-configures the nodes based on the user movements.}

\textcolor{correction2}{Our solution is a general approach to the node to x-area association, where x can be a handover region, a \ac{TA} or a set of TAs. Therefore, closely related to the question addressed by our solution is the autonomous configuration of TAs and TA Lists. Several solutions aiming at minimizing the signaling overhead caused by the periodic TA updates and paging have been developed in the research literature \cite{aqeeli2018dynamic} or patented \cite{wang2014method, singh2014tracking}. The approaches in \cite{aqeeli2018dynamic, wang2014method, singh2014tracking} are centralized and require information that does not exist in the network (e.g.~the \ac{UE} average speed, the average number of \ac{UE} per cell). Additionally, the authors do not provide implementation details in terms of data collection, and reconfiguration of the nodes. In contrast, our approach relies only on  information available in the network allowing an easy integration with the existing 4G and future 5G systems. }

\section{Handover Region Optimization}\label{sec:solution}

It is possible to minimize the number of inter-region handovers by dynamically re-configuring the association between \acp{BS} and nodes in the core, i.e. by grouping the \acp{BS} that exchange more users into the same handover region. In this Section we present two alternatives to the handover region optimization problem. We first start with a centralized optimal formulation. Due to the delay and the computational complexity of the \ac{CGP} approach, we also propose a distributed approach for the optimization. Finally, we show that the proposed distributed solution can be implemented by leveraging information already collected and exchanged within the network. While the centralized mechanism also relies on the same information, current standards do not require that information to be forwarded to the \ac{OSS}. This means that the centralized mechanism would be implemented as a passive probe in the core network. Hence, the optimal centralized solution could be deployed after the conditions in the network (\ac{UE} traffic pattern) have changed.

\begin{figure*}[h]
\centering
\subfloat[]{\includegraphics[trim=70 0 50 50,clip,scale=0.4]{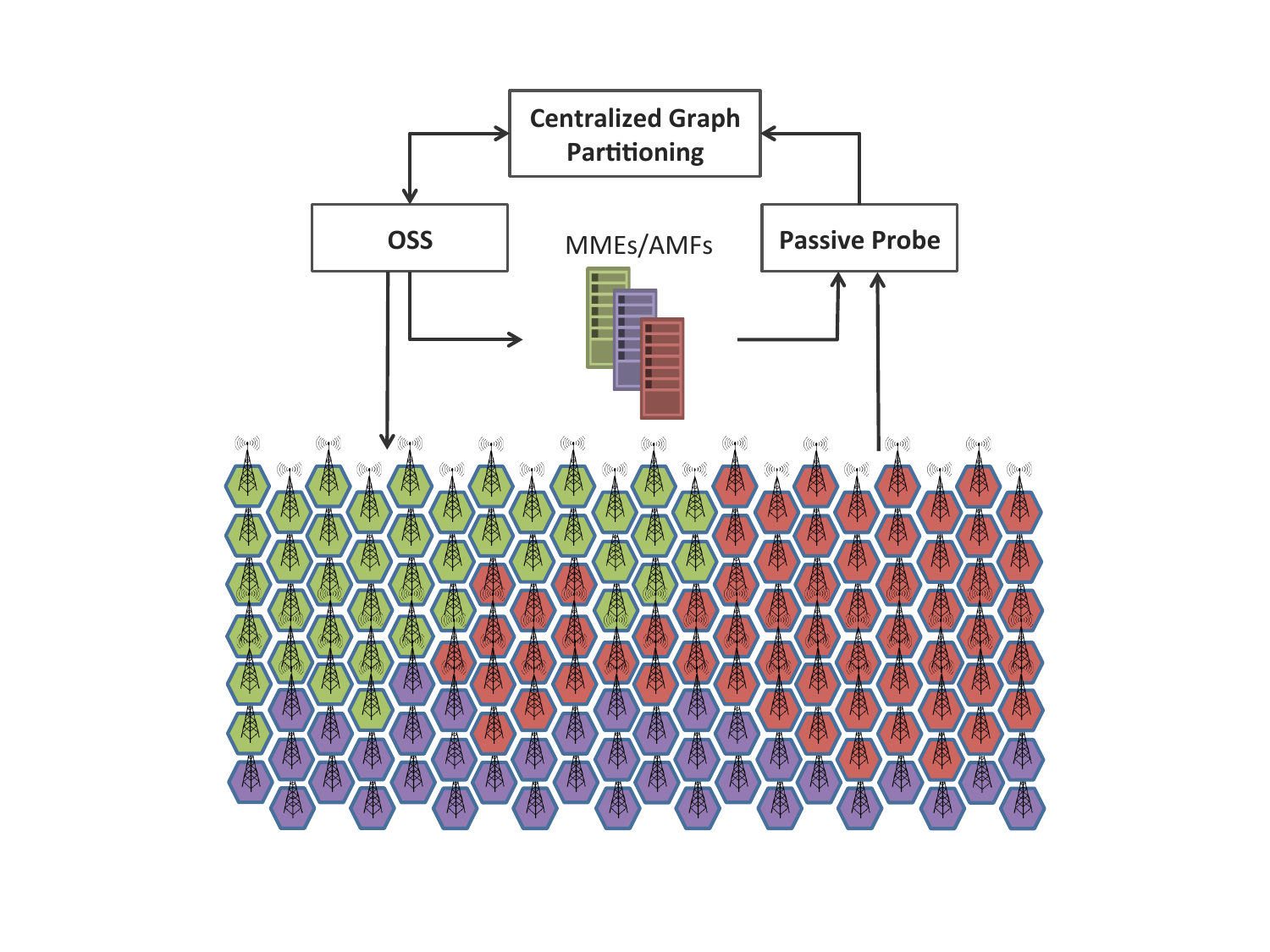}%
\label{fig:cgp_model}}
\hfil
\subfloat[]{\includegraphics[trim=70 0 50 50,clip,scale=0.4]{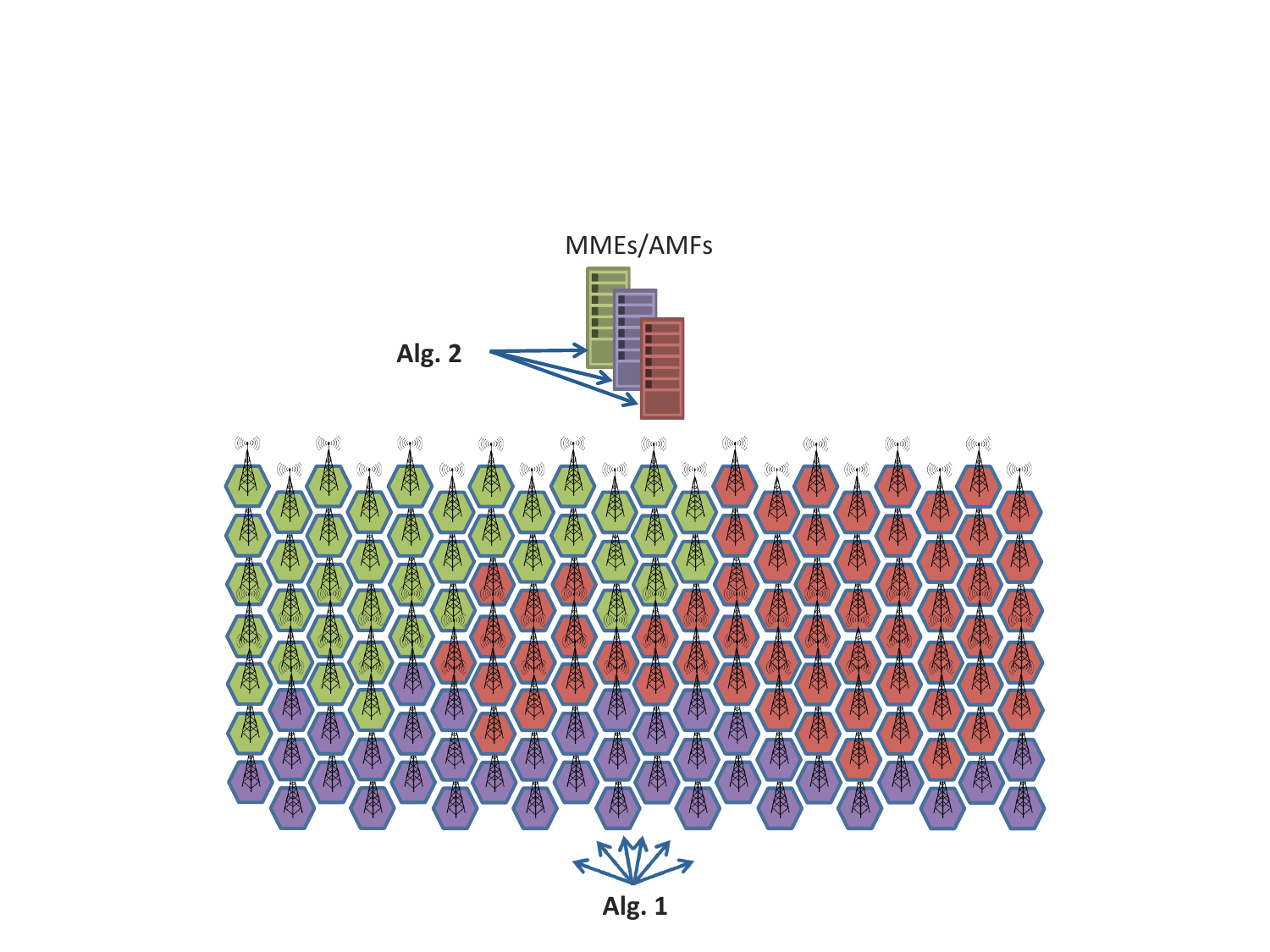}%
\label{fig:dso_model}}
\caption{\textcolor{correction2}{The two proposed alternatives for determining the  association of \acp{BS} to different \ac{MME}/\ac{AMF} (represented with different colors). (a) The centralized solution relies on a passive probe to collect the information about handovers while reconfiguring the nodes through the OSS. (b) With the distributed mechanism the nodes configuration is determined by each \ac{BS} and \ac{MME}/\ac{AMF} independently. In particular each \ac{BS} runs an instance of Alg. \ref{alg:pseudo_cell} and each \ac{MME}/\ac{AMF} runs an instance of Alg. \ref{alg:pseudo_mme}, while relying only on local information.   }  }
\label{fig:model}
\end{figure*}

\subsection{Centralized Graph Partitioning Approach}\label{sec:centralized_optimization_algorithm}


Re-configuring  \ac{RAN}-to-core nodes association can be formulated as a $k$-way \ac{GPP} on an undirected weighted graph $G = (V,E,w)$ that represents \acp{BS} as the set of vertices $V$ and the occurrence of handovers between \acp{BS} as the set of edges  $E \subseteq V \times V$. The weight $w_{i,j}$ is the normalized number of handovers that occurred  between nodes $i$ and $j$:
\begin{equation}\label{eq:weights_definition}
    w_{i,j} = \dfrac{h_{i,j} - h_{min}}{h_{max} - h_{min}},
\end{equation}
where $h_{i,j}$ is the number of handovers between nodes $i$ and $j$, $h_{max}$ and $h_{min}$ are the maximum and minimum  number of handovers that occurred between any two nodes in the graph. \textcolor{correction1}{For each edge $\{i,j\} \in E$ let us introduce a binary variable $z_{i,j}$, which is equal to $1$ iff $i$ and $j$ belong to two handover regions, i.e. to two \acp{MME}/\acp{AMF}. We also define a binary variable $x_{i,k}$ for each vertex $i\in V$ and each region $k \in K$, which is the set of all available \acp{MME}/\acp{AMF}. $x_{i,k}$ is equal to $1$ iff node $i$ belongs to handover region $k$. We denote the traffic load of each vertex $i\in V$ with $l_{i}$ and the load threshold of region $k\in K$ with $L_{k_{th}}$. The \ac{RAN}-to-core nodes association can be formulated as an integer linear programming problem:}


\begin{align}\label{eq:optimization_gcp}
\text{minimize}&\displaystyle\sum\limits_{\{i,j\} \in E} w_{i,j}z_{i,j} \nonumber\\
  \text{subject to}&\displaystyle\sum\limits_{k \in K}x_{i,k} = 1   ,   &\forall i \in V \nonumber\\
                 &\displaystyle\sum\limits_{i \in V}l_{i}x_{i,k} \leq L_{k_{th}}    ,    &\forall k \in K \\
                 & x_{i,k} - x_{j,k} \leq z_{i,j}   ,   & \forall{\{i,j\}} \in E, \forall k \in K \nonumber \\
                 & x_{j,k} - x_{i,k} \leq z_{i,j}   ,   & \forall{\{i,j\}} \in E, \forall k \in K \nonumber \\
                 & z_{i,j}\in\{0,1\} & \forall{\{i,j\}} \in E \nonumber \\
                 & x_{i,k}\in\{0,1\} & \forall{i} \in V, \forall k \in K \nonumber
\end{align}
Due to the NP-hard nature of the $k$-way \ac{GPP}, we rely on the \textsc{metis} library to solve \eqref{eq:optimization_gcp}\footnote{http://glaros.dtc.umn.edu/gkhome/views/metis}.
    \textcolor{correction2}{The centralized approach involves multiple stages: handover data collection at the \acp{BS};
    transferring the data to a centralized node in the core network;
    running the optimization; and transferring signaling messages in order
    to perform the re-configuration of the network (see Figure \ref{fig:cgp_model}). Hence, this approach results in an increased delay related to the network re-configuration.}




\subsection{Distributed Self-Organization
  Approach}\label{sec:distributed_self_organization_algorithm}

\begin{algorithm}[h]
\caption{RAN node optimization procedure}\label{alg:pseudo_cell}
\begin{algorithmic}[0] 
  \State $counters \gets \textit{\#handovers, \#handovers from each region} $
  \While {$\textit{Cell is operational}$}
  \State $\textit{Wait for event \{handover, reassign request\}} $
  \If {$\textit{event == handover}$}
  \State $\textit{Update counters}$
  \State $\textit{Calculate energy of attraction}$
  \State $\textit{MakeAssignmentDecision()}$
  \EndIf
  \If {$\textit{event == reassign request}$}
  \State $\textit{MakeReassignment(MME/AMF\_id)}$
  \EndIf
  \EndWhile
  \\\hrulefill
  \Function{$\textit{MakeAssignmentDecision()}$}{}
  \State $M \gets \textit{List of available MMEs/AMFs}$
  \State $k \gets \textit{number of MMEs/AMFs that are managing the cell}$
  \State $A \gets \textit{List of energies of attraction towards MMEs/AMFs}$
  \If {\textit{max(A) - current(A)$>$ ping pong threshold}}
  \State \textit{Send request to max(A) MME/AMF}
  \State \textit{Reset counters}
  \EndIf
  \EndFunction
  \\\hrulefill
  \Function{$\textit{MakeReassignment(MME/AMF\_id)}$}{}
  \State $A \gets \textit{List of energies of attraction towards MMEs/AMFs}$
  \State $\textit{Exclude the MME/AMF with MME/AMF\_id from A}$
  \State $\textit{Sort A in descending order}$
  \While {$\textit{MME/AMF not assigned}$}
  \State $\textit{Try to get assigned to an MME/AMF from A}$
  \State $\textit{Reset counters}$
  \EndWhile
  \EndFunction
\end{algorithmic}
\end{algorithm}

We now present a \ac{DSO} solution to the problem of handover region optimization. The proposed approach consists of two main components, one 
running on the \ac{RAN} nodes and the other one on virtual instances of core nodes (e.g. \acp{MME}/\acp{AMF}) (see Figure \ref{fig:dso_model}). \textcolor{correction2}{The computational complexity of both components is $O(1)$}.

The component running on the \ac{RAN} nodes is formalized in Alg.
\ref{alg:pseudo_cell}. 
It relies on  handover counters already available at the \acp{BS} (e.g. number of handovers, source handover \ac{MME}). The initial assignment of a \ac{BS} is chosen based on the majority of its neighbors. The optimization process is triggered whenever a handover occurs or in case an \ac{MME}/\ac{AMF} sends a reassignment request. In case of a handover, the \ac{BS} updates its counters and the algorithm calculates the energy of attraction towards all available \acp{MME}/\acp{AMF}. The \textbf{energy of attraction} of node $n$ towards the $m$-th \ac{MME}/\ac{AMF} is calculated as:
\begin{align}\label{eq:energy_of_attraction}
	A_m(n) = H_n(m)/\displaystyle\sum_{i \in K} H_n(i)
\end{align}
where $H_n(m)$ is the number of handover requests that arrived at node $n$ from nodes that are
assigned to the $m$-th \ac{MME}/\ac{AMF}. Therefore, the energy of
attraction towards an \ac{MME}/\ac{AMF} is the ratio between the number of handover
requests originating from this \ac{MME}/\ac{AMF} and the total number of handover requests
that arrived on the observed node.
Once the counters are updated and the energy of attraction is
calculated, the \ac{BS} decides whether to change its \acp{MME}/\acp{AMF} assignment based on the energy of attraction. It should be noted that the \ac{DSO} enables overlapping handover regions, i.e. a \ac{BS} can be connected to multiple \acp{MME}/\acp{AMF}. The \textit{ping pong threshold} restricts the \ac{BS} from constantly
changing its assigned \acp{MME}/\acp{AMF}. Once a \ac{BS} changes its assignment,
the counters are reset to default values and the adaptation process
restarts. The reset to the default values can be a hard reset, which resets the values to zero, or a soft reset, that sets the values of the counters by using a moving average.
In case an \ac{MME}/\ac{AMF} requests a reassignment of the \ac{BS} to another
\ac{MME}/\ac{AMF}, the \ac{BS}  attempts to get assigned to the next best (based on the energy of attraction)
available \ac{MME}/\ac{AMF}.

\begin{algorithm}[h]
\caption{\ac{MME}/\ac{AMF} optimization procedure}\label{alg:pseudo_mme}
\begin{algorithmic}[0] 
  \State $N \gets \textit{List of cells assigned to the MME/AMF} $
  \State $L \gets \textit{Current MME/AMF load}$
  \State $L_{max} \gets \textit{Load limit}$
  \State $A \gets \textit{List of energies of attraction of the cells}$
  \State $\textit{Wait for assignment request from n}$
  \If {$L + L(n) < L_{max}$}
  \State $\textit{Assign cell to the current MME/AMF}$
  \Else
  \If {$A(n) > min(A) + \delta$}
  \If {$L - L_{min(A)} + L(n) < L_{max}$}
  \State $\textit{Assign cell to the current MME/AMF}$
  \State $\textit{Inform the cell with min(A) to get reassigned}$
  \State $\textit{Remove the cell with min(A) from N and A}$
  \EndIf
  \Else
  \State $\textit{Reject the request}$
  \EndIf
  \EndIf
\end{algorithmic}
\end{algorithm}


The second component  runs on a virtual instance of
the \ac{MME}/\ac{AMF} and is formalized in Alg.
\ref{alg:pseudo_mme}. The \ac{MME}/\ac{AMF}
waits for a request from a node that wants to get assigned to it. If the combined
 load of the \ac{MME}/\ac{AMF} and the load coming from the node 
requesting the assignment is lower than a threshold ($L_{max}$), the request is accepted. In case the total load is greater than the
threshold, the assignment can be accepted or rejected depending on the energy of
attraction of the cell that is requesting the assignment. In particular, the assignment request is accepted only if among the cells currently attached to the \ac{MME}/\ac{AMF} there exists one with lower energy of attraction which, if removed, would free up enough resources to manage the requesting cell. If multiple such cells exist, the one with the lowest energy of attraction is selected, informed
to change its assignment, and removed from the list of cells assigned to the current \ac{MME}/\ac{AMF}.


\subsection{Integration with Network Protocols}\label{sec:integration_with_network_protocols}

\begin{figure}[t]
	\centering
	\includegraphics[scale=0.55]{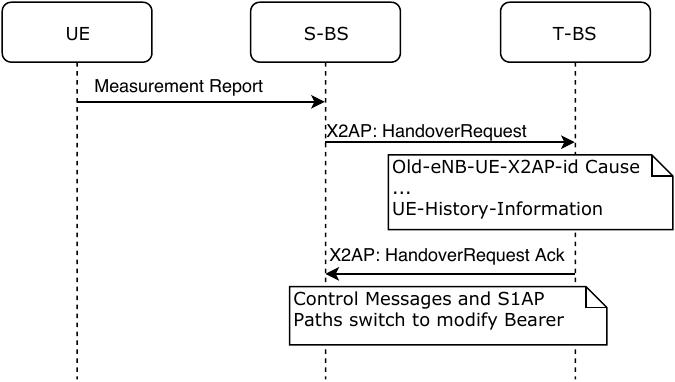}
	\caption{\textcolor{correction2}{Simplified sequence diagram of the intra-region handover procedure,
    with the emphasis on the \textit{HandoverRequest} message which carries the
    \textit{UE History Information}. S-BS and T-BS stand for Source and Target \ac{BS}.}}
	\label{fig:x2_handover_sequence_diagram}
\end{figure}

\textcolor{correction2}{It is important to highlight that the approaches described in Section \ref{sec:centralized_optimization_algorithm} and \ref{sec:distributed_self_organization_algorithm} have been designed to rely solely on information that is available through already existing signaling messages, removing the need for additional signaling and at the same time simplifying the integration with the existing architecture. The information of interest to the \ac{CGP} and \ac{DSO} is the type of handover, the number of handovers and the source \ac{MME} from which the handover originated. To explain this let us consider the procedures used to perform the two different types of handover (inter and intra-region). In 4G, the intra-region handovers are X2 and S1 without \ac{MME} reallocation. On the other hand, the inter-region handover procedure is the S1 with \ac{MME} reallocation. Similarly, in 5G, the equivalent of the intra-region handover procedures are Xn and N2 without \ac{AMF} reallocation; and the equivalent of the inter-region handover procedure is N2 with \ac{AMF} reallocation.}

\begin{figure}[t]
	\centering
	\includegraphics[scale=0.45]{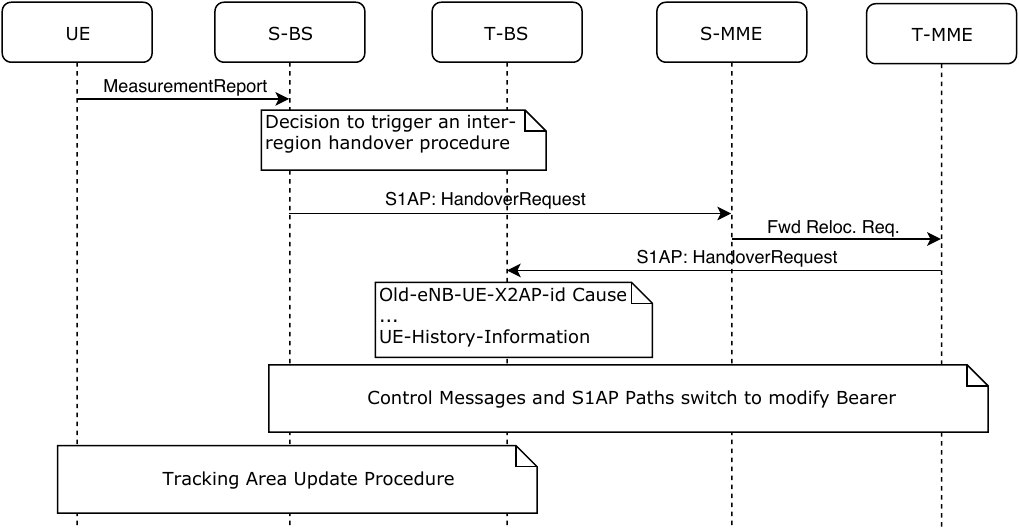}
	\caption{\textcolor{correction2}{Simplified sequence diagram of the inter-region handover procedure. The \textit{HandoverRequest} message  carries the
    \textit{UE History Information}, and the \ac{TAU} procedure
     is triggered after the handover.}}
	\label{fig:s1_handover_sequence_diagram}
\end{figure}

\begin{figure}[t]
	\centering
	\includegraphics[scale=0.55]{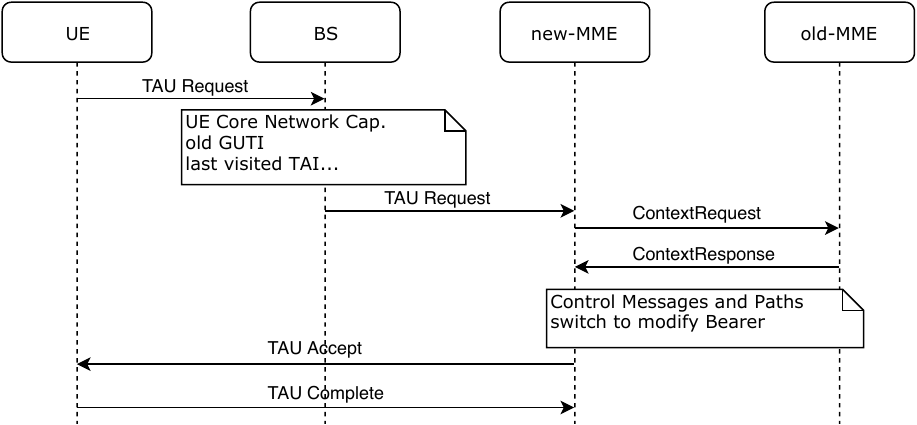}
	\caption{\textcolor{correction2}{Simplified sequence diagram of the \ac{TAU} procedure.
 The \textit{TAU Request} message carries the
    \textit{old GUTI} used to determine the source \ac{MME}.}}
	\label{fig:tau_procedure}
\end{figure}

\begin{figure*}[h]
\centering
\subfloat{\includegraphics[trim=200 160 0 0,clip,scale=0.16]{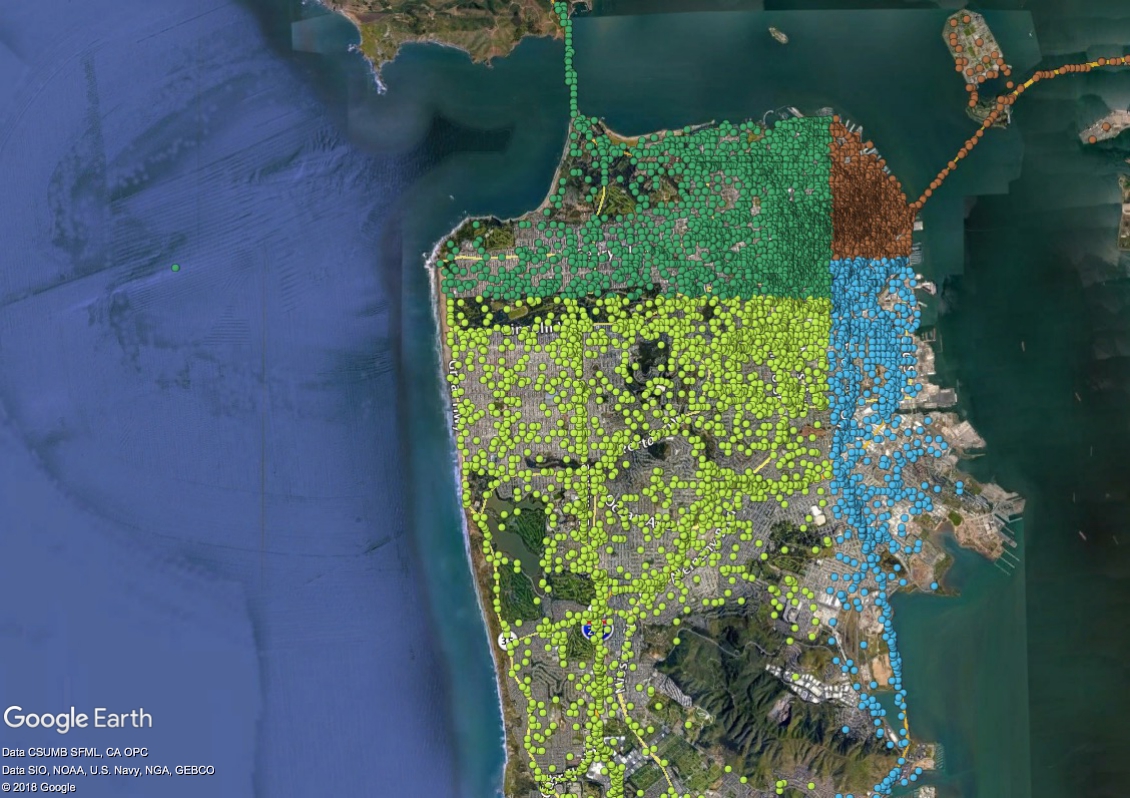}%
\label{fig:traditional_map}}
\hfil
\subfloat{\includegraphics[trim=200 160 0 0,clip,scale=0.16]{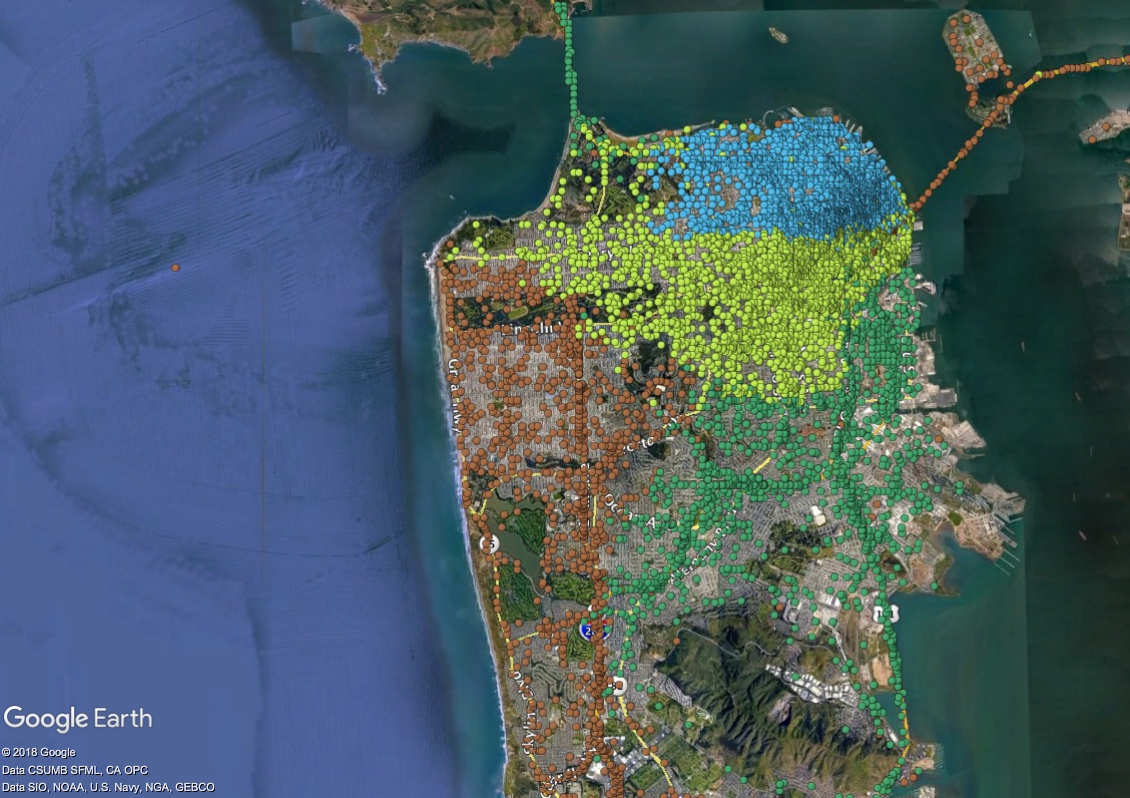}%
\label{fig:centralized_map}}
\hfil
\subfloat{\includegraphics[trim=200 160 0 0,clip,scale=0.16]{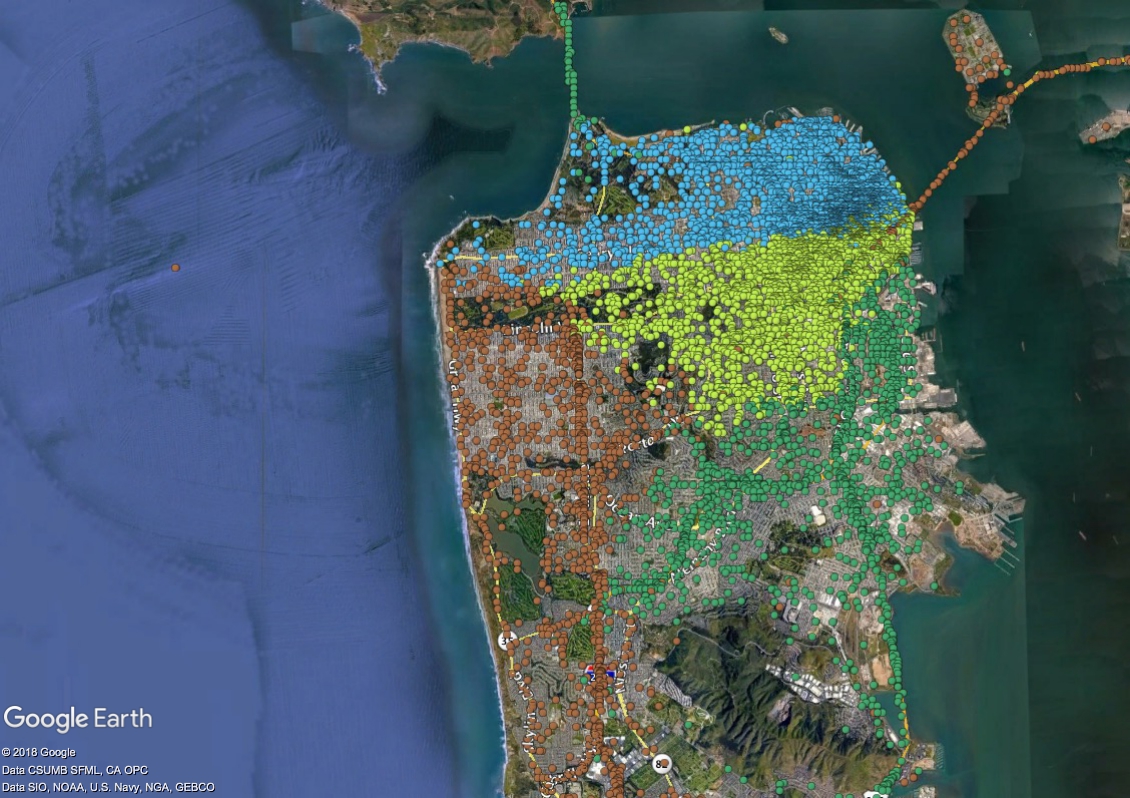}%
\label{fig:abm_map}}
\caption{Map of San Francisco area showing the \acp{BS} as colored dots; the grouping of \acp{BS} into handover regions is represented by different colors. From left to right the maps show the grouping based on the geographical approach, the grouping resulting from the \ac{CGP} based on data collected during a workday, and the grouping at 9PM of the same day resulting from the \ac{DSO}. For clarity of presentation we show only the \acp{BS} in the city.}
\label{fig:maps_for_the_traditional_centralized_and_abm_approach}
\end{figure*}

\textcolor{correction2}{As shown in Fig. \ref{fig:x2_handover_sequence_diagram}, the X2 handover procedure assumes direct communication between the involved nodes. The \textit{HandoverRequest} message contains a field reserved for the \textit{\ac{UE} History Information}, which contains information about the last visited cells by the \ac{UE} \cite{3gpp_2008_36_423}. This information is used to determine where the handover is coming from. Since the request is sent through the X2 link between two nodes, the handover type is obviously intra-region, i.e. the source \ac{MME} is equal to the destination \ac{MME}. This means that all counters can be updated appropriately (both the total number of handovers and the number of intra-region handovers are increased by $1$).}

\textcolor{correction2}{Fig. \ref{fig:s1_handover_sequence_diagram} shows the procedure that is used to perform a handover between nodes associated with different \acp{MME}. The initial \textit{HandoverRequest} message contains a field that transfers the information from the source to the target node (\textit{TargeteNB-ToSourceeNB-TransparentContainer}), which contains the \textit{\ac{UE} History Information} similarly to the previous case \cite{3gpp_36_413}. In case of an inter-region handover, we have to determine the source \ac{MME} as well in order to update all counters needed to calculate the energy of attraction in \eqref{eq:energy_of_attraction}. Considering that an \ac{MME} covers whole \acp{TA}, the inter-region handover includes a Tracking Area Update (TAU) as well (the last part of the handover procedure shown in Fig.  \ref{fig:s1_handover_sequence_diagram}), and in case of an intra-region S1 handover (an S1 handover without \ac{MME} reallocation) the \ac{TAU} is not required, which indicates to our algorithm that the source and target handover region are the same. As shown in Fig.  \ref{fig:tau_procedure}, the first message that is sent from the \ac{UE} to the \ac{BS} is the \textit{\ac{TAU} Request}, which contains information like \textit{\ac{UE} Core Network Capability, old \ac{GUTI}, last visited \ac{TA}}, etc. \cite{3gpp_2007_12}. The old \textit{\ac{GUTI}} is the identifier of interest to our algorithm, because it consists of two main components, namely the \textit{\ac{GUMMEI}} and the \textit{\ac{M-TMSI}}. Since the \textit{\ac{GUMMEI}} uniquely identifies the \ac{MME} which has allocated the \textit{\ac{GUTI}}, we have all the information needed to update all counters to calculate  \eqref{eq:energy_of_attraction}.}





\section{Performance evaluation}\label{sec:performance_evaluation}

In this section, we compare the performance of the \ac{DSO} approach detailed in Section \ref{sec:distributed_self_organization_algorithm}, the \ac{CGP} approach described in Section \ref{sec:centralized_optimization_algorithm}, and  the traditional static 
approach which groups  \acp{BS} into regions based on their location.
We rely on Agent Based Modeling, a computational model suitable for simulations of heterogeneous and self-organizing systems, to model the interactions between the \acp{BS} and the instances of \acp{MME}/\acp{AMF} whose behavior is formalized in Alg. \ref{alg:pseudo_cell} and \ref{alg:pseudo_mme}. For our simulations we draw on the dataset in \cite{OpenCellId}, which  provides \ac{BS} locations worldwide, and the dataset in \cite{SanFranTaxi}, which provides information about the movements of taxis in the San Francisco Bay Area, USA. We combine the two datasets to simulate the handover occurrences between \acp{BS}. In the considered area  $8,424$ \acp{BS} are active. In our simulations we followed the 3GPP recommendations \cite{3gpp_38_913_v_14_2_r_14} for the \ac{BS} coverage - $500$m coverage radius. We analyze the taxi movements for two days in order to capture the impact of the changes in user mobility patterns. \textcolor{correction2}{We assume that the load of each \ac{BS} is the same --- $l_i = 1, \forall i \in V$, and the load threshold of each region is $L_{k_{th}} = L_{max} =  \dfrac{|V|}{K}+1, \forall k \in K$. Hence, the partitioning defined with (\ref{eq:optimization_gcp}) results in an equal number of nodes in each partition. The ping point threshold in Alg. \ref{alg:pseudo_cell} is $0.25$, while $\delta=0$ in Alg. \ref{alg:pseudo_mme}.} 

Fig. \ref{fig:maps_for_the_traditional_centralized_and_abm_approach} zooms in on the grouping of \acp{BS} into handover regions in San Francisco. The \acp{BS} are grouped into four regions shown in green, blue, yellow and brown. Each region contains approximately the same number of \acp{BS}. The map on the left in Fig. \ref{fig:maps_for_the_traditional_centralized_and_abm_approach} shows the geographical approach to group the \acp{BS} into four regions with a balanced number of \acp{BS}. The map in the middle shows the grouping of the \acp{BS} resulting from the \ac{CGP}. Instead of relying on geographical proximity of \acp{BS}, the \ac{RAN}-to-core node association in this case is performed based on the user movement information collected during Day $1$. The map on the right in Fig. \ref{fig:maps_for_the_traditional_centralized_and_abm_approach} shows the \ac{RAN}-to-core node associations at 9PM of the same day resulting from the \ac{DSO} approach. The \ac{DSO} approach adapts over time: the \acp{BS} constantly monitor their counters and make decisions about their  association. The difference between the three maps is conspicuous, and it shows that the spatial proximity of \acp{BS} does not necessarily lead to a large exchange of users between them. For example, nearby \acp{BS} do not exchange a lot of users if their coverage areas overlap or if due to the terrain configuration it is impossible to move between them.

\begin{figure}[t]
	\centering
	\includegraphics[trim=10 20 10 70,clip,scale=0.24]{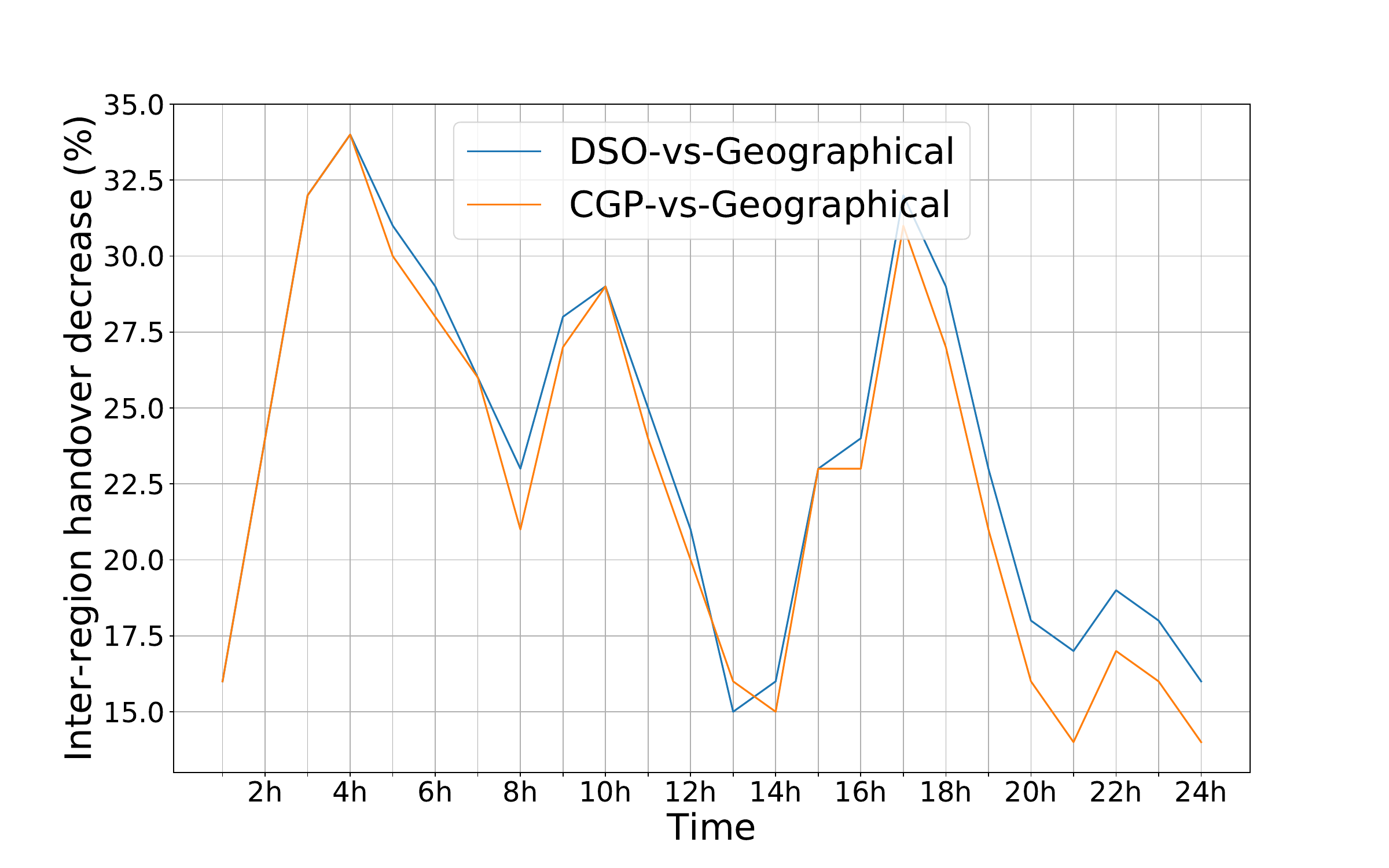}
	\caption{The hourly decrease of inter-region handovers during Day $1$; the \ac{CGP} approach relies on data for the whole $24$ hours of Day $1$. }
	\label{fig:inter_region_handover_decrease_workday}
\end{figure}

The map shown in the middle in Fig. \ref{fig:maps_for_the_traditional_centralized_and_abm_approach} is the ex post
optimal static association (\ac{CGP}) of nodes for the whole Day $1$ under consideration, i.e. the information over the whole Day $1$ is gathered and the best possible configuration for that period is computed. This might not be the best association at every point in time during the day since the number of handovers between \acp{BS} changes over time, and the weights $w_{i,j}$ in the objective in \eqref{eq:optimization_gcp} are the normalized total number of handovers between \acp{BS} over a period of time (in this case Day $1$). Since the \ac{CGP} requires information to be centrally collected, although theoretically possible, it is unlikely that the \ac{CGP} could be run over a shorter period. Let us now take
a closer look at how the share of inter-region handovers changes over time. Fig. \ref{fig:inter_region_handover_decrease_workday} shows the percentage of inter-region handovers during the Day $1$ of the \ac{DSO} and \ac{CGP} (ex post) with respect to the geographical clustering. 
As shown in Fig. \ref{fig:inter_region_handover_decrease_workday}
the decrease of inter-region handovers in case of \ac{CGP} is at least $15\%$ and at some point goes up to $33$\%. Perhaps more importantly, the
\ac{DSO} either performs as well as \ac{CGP}, or it outperforms it. This is due to its adaptive nature, which acts based on locally available information making the readjustments agile. 

\begin{table}[t]
\centering
\caption{Inter-region handover reduction in Day $1$ (blue) and $2$ (red).}
\label{table:gain_during_workday}
\begin{tabular}{l|c|c|cl}
  & Geographical  & \ac{CGP}  & \begin{tabular}[c]{@{}c@{}}\ac{DSO} \end{tabular} &  \\ \cline{1-4}
Geographical & \textcolor{blue}{$0$}/\textcolor{red}{$0$}& \textcolor{blue}{$-24,6\%$}/\textcolor{red}{-$20.9\%$}  & \textcolor{blue}{$-25.5$\%}/\textcolor{red}{$-23.7$\%}&  \\ \cline{1-4}
\ac{CGP} & \textcolor{blue}{$32.7$\%}/\textcolor{red}{$26.4$\%}& \textcolor{blue}{$0$}/\textcolor{red}{$0$}& \textcolor{blue}{$-1.2$\%}/\textcolor{red}{$-3.6$\%}&  \\ \cline{1-4}
\ac{DSO} & \textcolor{blue}{34.2\%}/\textcolor{red}{$31.1$\%}& \textcolor{blue}{$1.2$\%}/\textcolor{red}{$3.7\%$}& \textcolor{blue}{$0$}/\textcolor{red}{$0$}& 
\end{tabular}
\end{table}

\begin{figure}[t]
	\centering
	\includegraphics[trim=10 20 10 70,clip, scale=0.24]{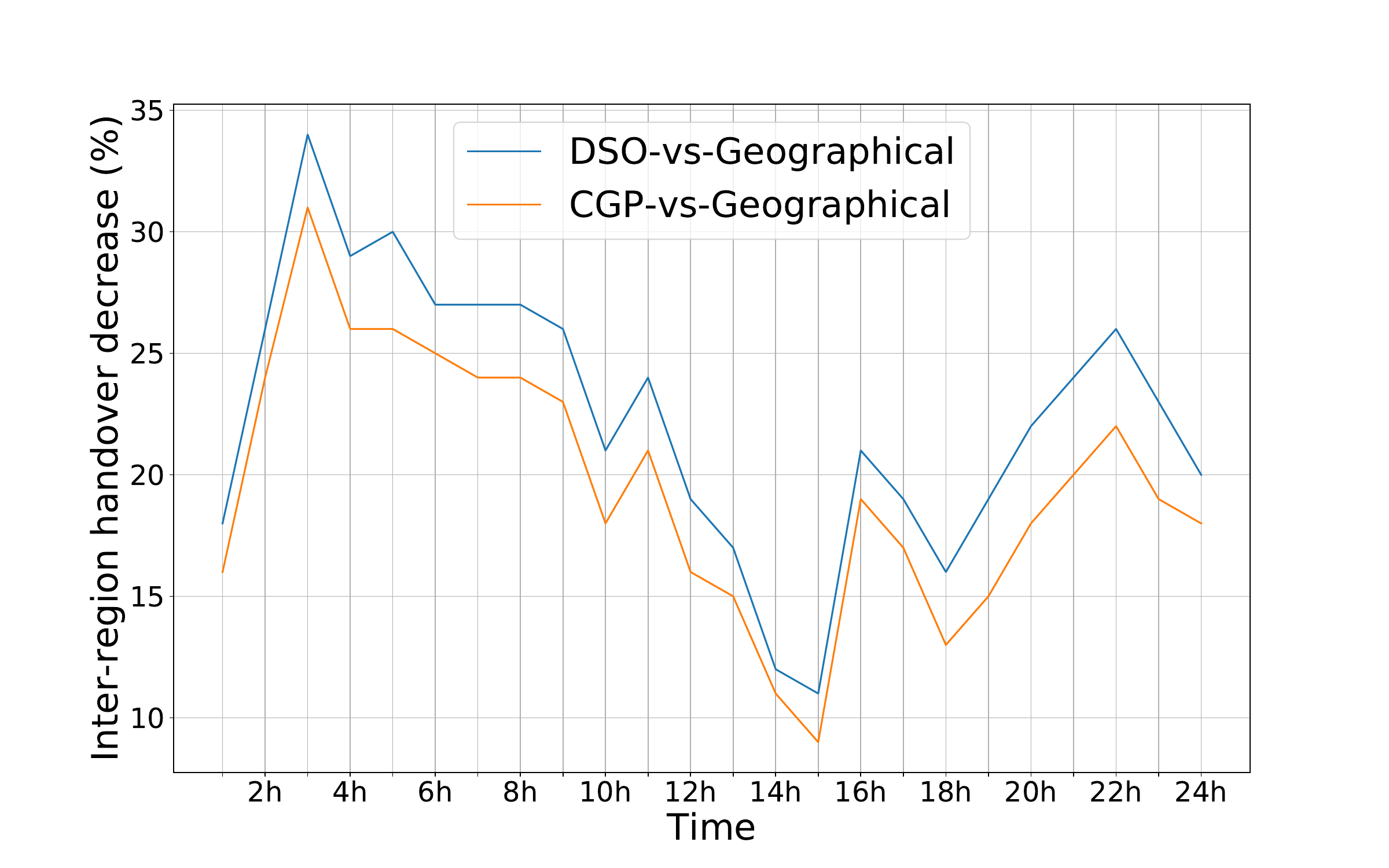}
	\caption{Hourly decrease of inter-region handovers during Day $2$; the \ac{CGP} configuration evaluated in this plot results from Day $1$ mobility data to emulate the delay resulting from collecting and processing mobility data at the \ac{OSS}.}
	\label{fig:inter_region_handovr_decrease_weekend}
\end{figure}

Fig. \ref{fig:inter_region_handovr_decrease_weekend} shows the hourly decrease of inter-region handovers during Day $2$. The performance of the \ac{CGP} approach corresponds to the same \ac{RAN}-to-core node association used in Fig. \ref{fig:inter_region_handover_decrease_workday}. This is a more fair comparison between the \ac{CGP} and the \ac{DSO} approach, in that the \ac{RAN}-to-core node association is computed based on historical data and then used in the network. In this case the gain achieved with the \ac{CGP} approach is evidently lower compared to the \ac{DSO} approach. Both approaches - even the \ac{CGP} one relying on outdated information - significantly outperform the geographical association.

Now, let us examine the overall impact on the number of inter-region handovers in the network. Table \ref{table:gain_during_workday} shows  the  reduction of inter-region handovers for Day $1$ (blue) and Day $2$ (red). Both  the \ac{CGP} and \ac{DSO} perform significantly better compared to the geographical approach, with a minimum gain of $26.4\%$. 
The gain of the \ac{DSO} approach compared to the \ac{CGP} one is $3.7\%$ in Day $2$, which is more than two times higher compared to the gain during Day $1$. This is due to changes in the user movement patterns: a static approach to optimizing the \ac{RAN}-to-core nodes association is suboptimal. 

\begin{figure}[t]
	\centering
	\includegraphics[scale=0.45]{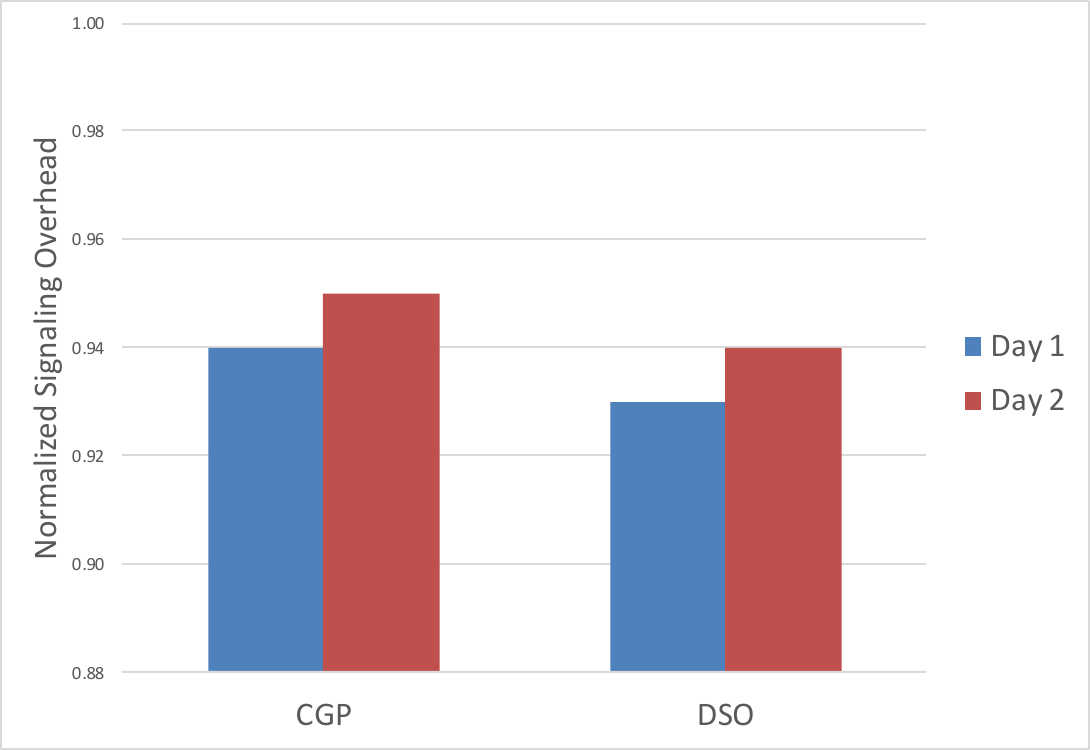}
	\caption{\textcolor{correction2}{Normalized signaling overhead related to the execution of mobility management functions. The values are normalized by the signaling overhead that results from the geographical approach to RAN-to-core node association. For the handover events in our network deployment, a $5$\% reduction is equal to $1$ billion signaling messages.}}
	\label{fig:num_sig_msg_overall}
\end{figure}

\textcolor{correction2}{We also study the signaling overhead related to the handover procedures. As previously mentioned in Section \ref{sec:introduction}, inter-region handovers require a significantly larger number of signaling messages compared to intra-region handovers. Therefore, by minimizing the number of inter-region handovers, we minimize the overall network signaling overhead. Figure~\ref{fig:num_sig_msg_overall} compares the normalized signaling overhead related to the execution of handovers for the \ac{CGP} and \ac{DSO} approaches. As shown in Figure~\ref{fig:num_sig_msg_overall}, both optimization techniques, i.e.~the \ac{CGP} and the \ac{DSO}, clearly outperform and minimize the signaling overhead compared to the traditional geographical clustering approach.}

\begin{figure}[t]
	\centering
	\includegraphics[scale=0.45]{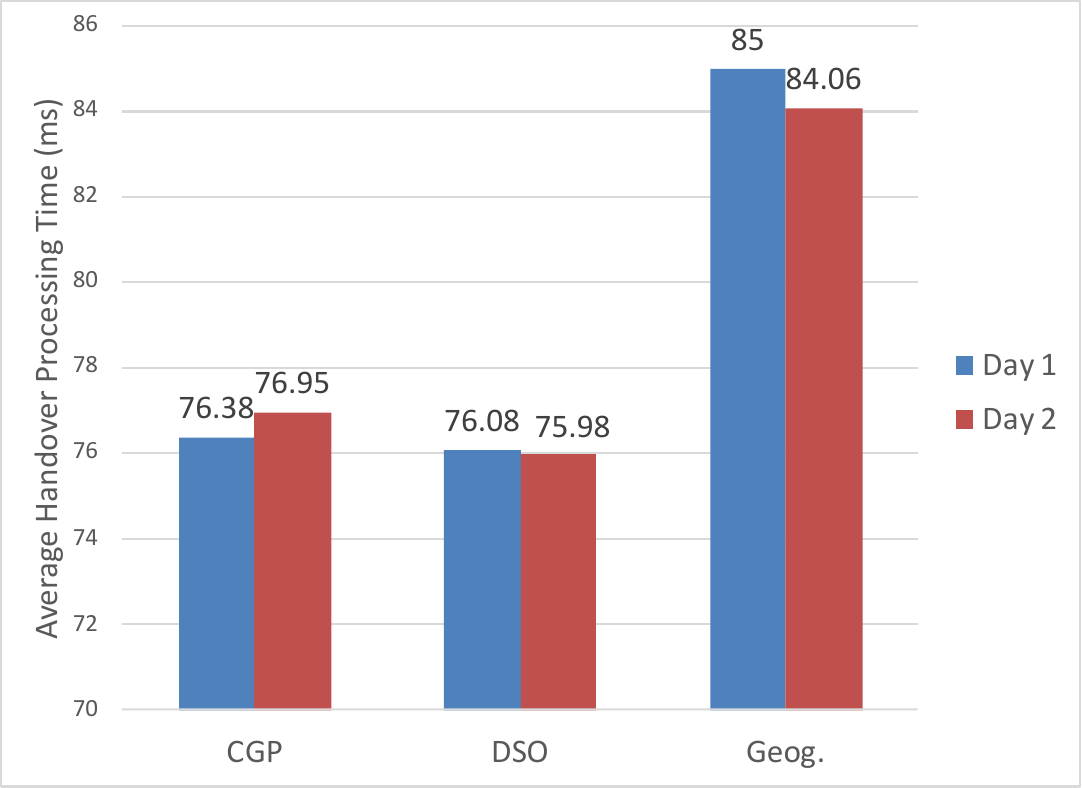}
	\caption{\textcolor{correction2}{Average handover procedure processing time (ms).}}
	\label{fig:signaling_processing_time_overall}
\end{figure}

\textcolor{correction2}{The reduction related to the signaling overhead affects the average handover procedure processing time. According to \cite{3gpp_25_912v13,li2010user} the average handover procedure processing time also depends on the type of handover. The processing of intra-region handovers is estimated to $50$ms, whereas the processing time of inter-region handovers is estimated to be in the range from $100$ms to $350$ms. Figure~\ref{fig:signaling_processing_time_overall} shows the comparison between the average handover procedure processing time for the \ac{CGP}, \ac{DSO} and geographical approach. As shown in Figure~\ref{fig:signaling_processing_time_overall}, the \ac{CGP} and the \ac{DSO} algorithms significantly reduce the average handover procedure processing time. The results presented in Figure~\ref{fig:num_sig_msg_overall} and Figure~\ref{fig:signaling_processing_time_overall} clearly highlight the benefits from the handover region optimization techniques presented in this paper. They also prove that the delay and signaling overhead related to handover procedures can be reduced without the need to change the procedures themselves, but rather by reconfiguring the association between the \ac{RAN} nodes and the nodes in the core network.}

\section{Conclusions}\label{sec:conclusions}

 We proposed a method to reorganize nodes in the network so that the number of handovers requiring an \ac{MME}/\ac{AMF} reallocation is minimized. We presented a distributed (\ac{DSO}) and a centralized approach (\ac{CGP}). \textcolor{correction2}{In the case of the centralized approach, the novelty consists in formalizing the handover regions optimization as a graph partitioning optimization problem that only relies on information already used within 4G and 5G networks. In the case of the distributed approach, the additional novelty consists in proposing a new method that optimizes the handover regions in a distributed manner and only relies on local information already available to each node (base stations and MME/AMF) in a 4G and 5G network.
 Both approaches outperform the traditional geographic clustering of \acp{BS} and significantly reduce the average handover procedure processing time. We validated the proposed approaches using real user mobility datasets.}
 By relying on distributed decision making, the \ac{DSO} can adapt to the changes in the user movement patterns as they happen, and outperforms \ac{CGP} removing also the need for collecting the information at the \ac{OSS}. Future work will focus on modelling overlapping handover regions in the centralized approach (which is currently only included in \ac{DSO}) and examine the implications of dynamic up- and down- scaling of \acp{MME}/\acp{AMF}.

\begin{acronym}
  \acro{3GPP}{3rd Generation Partnership Project}
  \acro{BSs}{Base Stations}
  \acro{BS}{Base Station}
  \acro{TAs}{Tracking Areas}
  \acro{TAL}{Tracking Area List}
  \acro{TA}{Tracking Area}
  \acro{ABM}{Agent-Based Modeling}
  \acro{EPC}{Evolved Packet Core}
  \acro{MME}{Mobility Management Entity}
  \acro{SGW}{Serving Gateway}
  \acro{PGW}{Packet Gateway}
  \acro{HSS}{Home Subscriber Server}
  \acro{M2M}{Machine to Machine}
  \acro{5G}{Fifth Generation}
  \acro{SDN}{Software Defined Networking}
  \acro{NFV}{Network Function Virtualization}
  \acro{VNF}{Virtualized Network Function}
  \acro{IoT}{Internet of Things}
  \acro{KPI}{Key Performance Indicator}
  \acro{GSM}{Global System for Mobile Communications}
  \acro{UMTS}{Universal Mobile Telecommunications System}
  \acro{LTE}{Long Term Evolution}
  \acro{C-RAN}{Cloud Radio Access Network}
  \acro{FAP}{Femtocell Access Point}
  \acro{UE}{User Equipment}
  \acro{3GPP}{3rd Generation Partnership Project}
  \acro{MB-ICIC}{Mobility-Based Inter-Cell Interference Coordination}
  \acro{MAC}{Medium Access Control}
  \acro{RACH}{Random-Access Channel}
  \acro{RAN}{Radio Access Network}
  \acro{PCRF}{Policy control and Charging Rules Function}
  \acro{TAU}{Tracking Area Update}
  \acro{SBA}{Service Based Architecture}
  \acro{AMF}{Access and Mobility Function}
  \acro{UPF}{User Plane Function}
  \acro{SMF}{Session Management Function}
  \acro{PCF}{Policy Control Function}
  \acro{GPP}{Graph Partitioning Problem}
  \acro{GCP}{Graph Coloring Problem}
  \acro{ILP}{Integer Linear Programming}
  \acro{CGP}{Centralized Graph Partitioning}
  \acro{DSO}{Distributed Self-Organized}
  \acro{OSS}{Operations Support System}
  \acro{GUTI}{Globally Unique Temporary UE Identity}
  \acro{GUMMEI}{Globally Unique MME Identifier}
  \acro{M-TMSI}{MME Temporary Mobile Subscriber Identity}
\end{acronym}


%

\bibliographystyle{IEEEtran}  
\bibliography{main}

\end{document}